\documentstyle[psfig,amsmath]{article}

\def\bd{{\bf d}}

\def\bk{{\bf k}}

\def\bp{\vec{p}}

\def\nd{\phantom{\dagger}}
\def\beq{\begin{equation}}
\def\ee{\end{equation}}
\def\bea{\begin{eqnarray}}
\def\eea{\end{eqnarray}}
\def\cH{{\cal H}}
\def\cT{{\cal T}}
\def\cU{{\cal U}}
\def\cD{{\cal D}}
\def\ij{{\langle i, j \rangle}}

\def\refq#1{(\ref{#1})}

\begin{document}

\title{Quantum Magnetism Approaches to Strongly Correlated Electrons}

\author{Assa Auerbach \\ 
{\em Physics Department, Technion, Haifa, Israel.} \\ \vspace{0.3cm} \\
Lecture notes compiled by \\ \vspace{0.01cm}
 \\  Federico Berruto \\ 
{\em Dipartimento di Fisica and INFN, Perugia, Italy}  
\\ and \\ Luca Capriotti \\ {\em 
International School for Advanced Studies (SISSA)} \\ 
{\em  and INFM, Trieste, Italy.} }

\date{Summer School at Chia Laguna , 31 August - 11 September 1997}

\maketitle

\begin{abstract}
Problems of strongly interacting electrons can be greatly simplified by
reducing them to effective quantum spin models. The initial step is
renormalization
of the Hamiltonian into a lower energy subspace.
The positive and negative U Hubbard models are explicitly
transformed into the Heisenberg
and -x-xz models respectively.
Basic tools of quantum magnetism are introduced and used: 
spin coherent states path integral, spin wave theory, and  continuum
theory of  rotators. 
The last lecture concerns pseudospin approaches to superconductivity and superfluidity.
The SO(3) rotator theory for the -x-xz model describes the charge density wave to superconductor transition
for e.g. doped bismuthates.
Analogously, Zhang's  theory for collective modes of high T$_c$ cuprates
describes the antiferromagnet to $d$-wave superconductor
transition using SO(5)  rotators. 
Finally,  the Magnus force on two dimensional vortices 
and their momentum, are derived from the Berry phase  of the spin path integral.
\end{abstract}  
\newpage 

\tableofcontents

\newpage

\part{Deriving the Effective Hamiltonian}
\label{lec1}
Let us consider the Hubbard model for
conduction electrons hopping on a lattice with short range interactions
\bea
\cH&=&\cT+\cU\nonumber\\
\cT &=&- t \sum_{\langle ij\rangle s=\uparrow,\downarrow} 
 c_{i,s}^{\dagger}c^{\nd}_{j,s}\nonumber\\
\cU &=& U\sum_{i} n_{i,\uparrow}n_{i,\downarrow},
\label{HM}
\end{eqnarray}
It is always tempting to reduce the interaction term $\cU$ to fermion bilinears 
(single electron
terms) using the 
Hartree Fock (HF)  variational approximation.  However this approach is
 known to be seriously flawed in several important
cases. For example, while the HF spin density wave is  energetically favorable   for 
$U>0$ at half filling,  
it breaks spin symmetry too readily in one and two dimensions, in violation of the Mermin Wagner theorem.
This implies that 
Fock states might be too restrictive as a variational basis. We can illustrate this point
using a simple toy model: 
the Hubbard model on two sites. It will also teach us 
something about onsite interactions and their effect on 
spin correlations. 
\section{Two-site Hubbard model}
\label{s.2siteHubbmod}
The two-site Hubbard model with two electrons,  is
\begin{equation}
H = - t \sum_{s=\uparrow,\downarrow} \left( c_{1,s}^{\dagger} c_{2,s}
+ c_{2,s}^{\dagger} c_{1,s} \right) 
+ U\sum_{i=1,2} n_{i,\uparrow} n_{i,\downarrow},
\end{equation}
where  we take $ U>0 $ .
The total spin ${\bf S} ={\bf S}_{1}+{\bf S}_{2} $, commutes with the 
Hamiltonian, we restrict ourselves to the $S=0$ and $S=1$ subspaces.
The singlet subspace is spanned by three states
\begin{equation}
\frac{ c_{1,\uparrow}^{\dagger}c_{2,\downarrow}^{\dagger} - 
                                     c_{1,\downarrow}^{\dagger}c_{2,\uparrow}^{\dagger}}
                                    {\sqrt{2} }  \left| 0 \right> ~~,~~
 c_{1,\uparrow}^{\dagger}c_{1,\downarrow}^{\dagger}  \left| 0 \right>            
~~,~~c_{2,\uparrow}^{\dagger}c_{2,\downarrow}^{\dagger} \left| 0 \right>~.
\end{equation}
In this subspace, the Hamiltonian is
\begin{displaymath}
H_{S=0}=
\left(\begin{array}{ccc}
0          & \sqrt{2}t & \sqrt{2}t \\
\sqrt{2}t  &    U      &     0     \\
\sqrt{2}t  &    0      &     U     \\
\end{array} \right)~.
\end{displaymath}
Diagonalizing the Hamiltonian in the singlet sector, one gets
\begin{displaymath}
H_{S=0} \longrightarrow 
\left(\begin{array}{ccc}
   \left(U-\sqrt{U^{2}+16t^2}\right)/2    &    0       &         0    \\
     0                                    &    U      &          0    \\
     0                                    &    0      &   \left(U+\sqrt{U^{2}+16t^2}\right)/2 \\
\end{array} \right)~,
\end{displaymath}
which, in the {\em strong coupling limit} ($U \gg t$) is
\begin{displaymath}
H_{S=0} \xrightarrow[U \gg t]{} 
\left(\begin{array}{ccc}
   -4t^{2}/U    &    0      &          0          \\
     0          &    U      &          0          \\
     0          &    0      &   U + 4t^{2}/U      \\
\end{array} \right)~.
\end{displaymath}
The triplet subspace is spanned by the  states:
\begin{equation}
\frac{ c_{1,\uparrow}^{\dagger}c_{2,\downarrow}^{\dagger} + 
                                     c_{1,\downarrow}^{\dagger}c_{2,\uparrow}^{\dagger}}{ \sqrt{2}}
                                     \left| 0 \right> ~~,~~
 c_{1,\uparrow}^{\dagger}c_{2,\uparrow}^{\dagger}\left| 0 \right>~~
,~~c_{1,\downarrow}^{\dagger}c_{2,\downarrow}^{\dagger}\left| 0 \right>  ~, 
\end{equation}
which all have zero energy.

\begin{figure}[t]
\begin{picture}(325,200)
\put (35,0){\vector(0,1){200}}
\put (15,190){\large E}
\put (15,100){  0}
\put (15,170){ $2t$}
\put (10,30){$-2t$}

\put (32.5,100){\line(1,0){2.5}}
\put (32.5,170){\line(1,0){2.5}}
\put (32.5,30){\line(1,0){2.5}}

\put (42,100){\line(1,0){30}}
\put (42,170){\line(1,0){30}}
\put (42,30){\line(1,0){30}}

\put (46,105){$1 \times S$}
\put (46,88){$ 3 \times T$}
\put (46,175){$1 \times S$}
\put (46,35){$1 \times S$}

\put (82,27.5){$
\frac{1}{2}
\Big[\big(\left|\uparrow_{1},\downarrow_{2}\right> 
- \left| \downarrow_{1},\uparrow_{2}\right>\big) 
+\big(\left|\uparrow\downarrow_{1},0_{2}\right>
+\left|0_{1},\uparrow\downarrow_{2}\right>\big)\Big] 
$ }
\put (82,167.5){$ 
\frac{1}{2}
\Big[\big(\left|\uparrow_{1},\downarrow_{2}\right> 
- \left| \downarrow_{1},\uparrow_{2}\right>\big) 
-\big(\left|\uparrow\downarrow_{1},0_{2}\right>
+\left|0_{1},\uparrow\downarrow_{2}\right>\big)\Big] 
$ }
\put (82,97.5){$\bigg\{ $}
 
\put (92,110){$ 
\frac{1}{\sqrt{2}}
\big(\left|\uparrow\downarrow_{1},0_{2}\right>
-\left|0_{1},\uparrow\downarrow_{2}\right>\big) 
$}
\put (92,85){$ 
\left|\uparrow_{1},\uparrow_{2}\right>~, \, \left|\downarrow_{1},\downarrow_{2}\right>~, \,
\frac{1}{\sqrt{2}}
\big(\left|\uparrow_{1},\downarrow_{2}\right>
+\left|\downarrow_{1},\uparrow_{2}\right>\big) 
$}

\end{picture}
\caption{Noninteracting ($U$=0) eigenstates of  the two-site 
Hubbard model. The ground state is the  ``two electron Fermi sea''.}
\label{hubbun}
\end{figure}
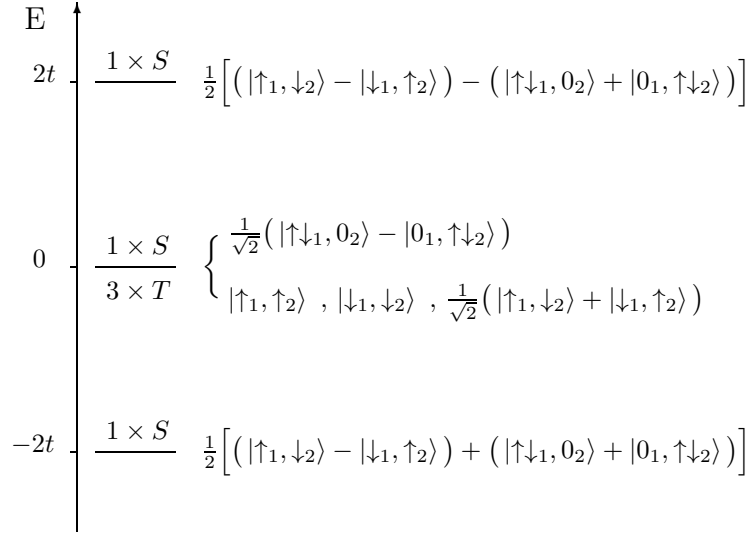

\begin{figure}[t]
\begin{picture}(325,200)
\put (35,0){\vector(0,1){200}}
\put (10,190){\large E}
\put (15,70){  0}
\put (15,170){$U$}
\put (-2,30){$-4t^2/U$}

\put (32.5,70){\line(1,0){2.5}}
\put (32.5,170){\line(1,0){2.5}}
\put (32.5,30){\line(1,0){2.5}}

\put (42,70){\line(1,0){30}}
\put (42,170){\line(1,0){30}}
\put (42,185){\line(1,0){30}}
\put (42,30){\line(1,0){30}}

\put (46,75){$3 \times T$}
\put (46,175){$1 \times S$}
\put (46,190){$1 \times S$}
\put (46,35){$1\times S$}

\put (82,165.5){$ 
\frac{1}{\sqrt{2}}
\big(\left|\uparrow\downarrow_{1},0_{2}\right>
-\left|0_{1},\uparrow\downarrow_{2}\right>\big) 
$}
\put (82,185.5){$ 
\frac{1}{\sqrt{2}}
\big(\left|\uparrow\downarrow_{1},0_{2}\right>
+ \left|0_{1},\uparrow\downarrow_{2}\right>\big) 
$}

\put (82,67.5){$ 
\left|\uparrow_{1},\uparrow_{2}\right>~, \, \left|\downarrow_{1},\downarrow_{2}\right>~, \,
\frac{1}{\sqrt{2}}
\big(\left|\uparrow_{1},\downarrow_{2}\right>
+\left|\downarrow_{1},\uparrow_{2}\right>\big) 
$}

\put (82,27.5){$
\frac{1}{\sqrt{2}}
\big(\left|\uparrow_{1},\downarrow_{2}\right>
-\left|\downarrow_{1},\uparrow_{2}\right>\big) 
$}

\end{picture}
\caption{Eigenstates of the two-sites Hubbard model in the
strong coupling limit.}
\label{hubbper}
\end{figure}
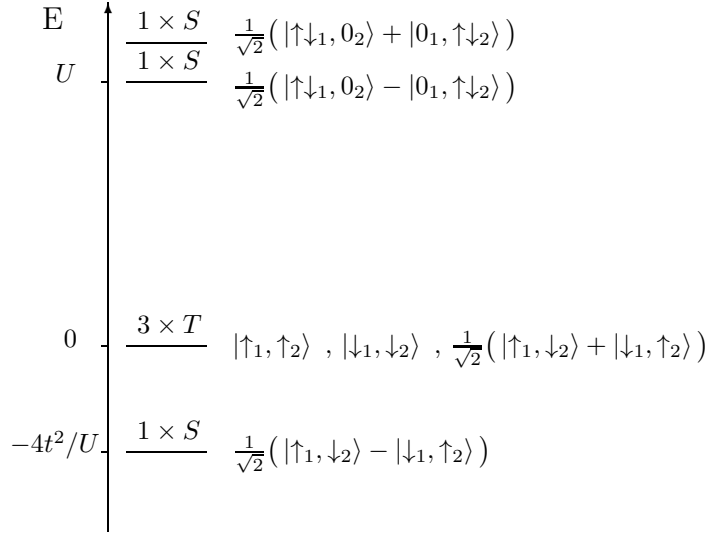

The comparison of the spectrum and wave functions of  the non-interacting case and 
the strong coupling limit is displayed in
Figs.~\ref{hubbun} and \ref{hubbper}.

The non-interacting groundstate is the two-site version
of the Hartree-Fock two electron ``Fermi sea''.
This state, and the lowest excited multiplet, contain sizeable contributions from the
doubly occupied
singlets
\begin{equation}
\frac{1}{\sqrt{2}}\left( \left|\uparrow\downarrow_{1},0_{2} \right> 
,~~~                              \left|0_{1},\uparrow\downarrow_{2} \right>\right) 
.
\end{equation}
The Hubbard interaction pushes these states to energies 
of order $U$.  At strong coupling therefore,  the ground state becomes a {\em valence bond singlet},
of singly occupied sites with no
charge fluctuations. It also {\em cannot be expressed, even approximately, as a Fock state.}  

The lessons  to be learned from this toy model is that repulsive
interactions can
\begin{itemize}
\item  enhance magnetic correlations,
\item  reduce double occupancies in the ground state,
\item  separate spin and charge excitations.
\end{itemize}

\section{Renormalization to Low Energy Subspace}
\label{s.lowenproj}

Low temperature phases and their interesting 
DC transport properties, are determined by the 
ground state 
and low energy excitations. This chapter is slightly formal, as it shows what is precisely meant by the
Renormalization Group  (RG) transformation which replaces an orginal non diagonal Hamiltonian with an effective 
one for a lower energy subspace.
Let us consider any Hamiltonian written as
\begin{equation}
H = H_{0} + V ~,
\end{equation}
where $H_{0}$ is diagonal and $V$ is a non diagonal perturbation.
We define the Hilbert space using the eigenstates of $H_{0}$, and $P_{0}(\Lambda)$ is projector onto
the subspace with energies less than $\Lambda$. 

The resolvent operator, $G = (E - H)^{-1}$, projected onto the latter 
subspace is given by a well known matrix inversion identity\cite{CRC}
\begin{multline}
G_{00}(E) = 
P_{0}G(E)P_{0} = \\  = P_{0}\left(\begin{array}{cc}
E - P_{0}(H_{0}+V)P_{0}     &    P_{0}V(1-P_{0})       \\
(1-P_{0})VP_{0}             &    (1-P_{0})(E-(H_{0}+V))(1-P_{0}) \\
\end{array}\right)^{-1}P_{0} = \\
 = \big\{E - P_{0}(H_{0}+V)P_{0} - \\
-  P_{0}V(1-P_{0})\big[(1-P_{0})(E-(H_{0}+V))(1-P_{0})\big]^{-1}(1-P_{0})VP_{0} \big\}^{-1}  \\ 
\equiv  \left(E-{\mathcal H}_{eff}(E)\right)^{-1}~,
\label{e.effham}
\end{multline}
where the last equality defines the {\em effective Hamiltonian} ${\mathcal H}_{eff}(E)$.
Then the spectrum of $H$, which corresponds to states with non-zero
weights 
in the subspace considered, is given by the zeros of the characteristic polynomial of
${\mathcal H}_{eff}(E)$,
\begin{equation}
\det\left(E_{n} - {\mathcal H}_{eff}(E_{n})\right)
\end{equation}
that is by the poles of the function $\textrm{Tr}\,G(E)_{00}$.
The effective Hamiltonian can be also written as
\begin{multline}
{\cal H}_{eff}(E) =  P_{0}\bigg\{ (H_{0}+V) + V(1-P_{0})\times \\
\times\bigg[(1-P_{0})\bigg(1-(E-H_{0})^{-1}V(1-P_{0})\bigg)\bigg]^{-1}(1-P_{0})(E-H_{0})^{-1}V\bigg\}P_{0}= \\
= P_{0}\bigg\{(H_{0}+V) + V \sum_{n=1}^{\infty}\bigg[\frac{1-P_{0}}{E-H_{0}}V\bigg]^{n}\bigg\}P_{0}~.
\label{effexp}
\end{multline}
If $P_0$ projects onto the  ground state manifold of $H_0$, Eq. (\ref{effexp})
defines the {\em Brillouin-Wigner} 
perturbation theory\footnote{Note that the sum $\sum_{n=1}^{\infty}$  in  Eq. (\ref{effexp}) does not
correspond
to a perturbation series for the ground state energy, since the terms depend   on $E$.}.
For two cut-off energies   $\Lambda ' < \Lambda$, Eq.~\refq{effexp} is a 
Renormalization Group transformation
\beq
\cH(\Lambda) \to \cH(\Lambda') = H_0(\Lambda') + V(\Lambda')
\end{equation}
where
\begin{equation}
{\cal H} (\Lambda') = P(\Lambda')\bigg\{{\cal H} (\Lambda ) 
- V(\Lambda )\sum_{n=1}^{\infty}\bigg[\frac{1-P({\Lambda'})}{E-H_{0}(\Lambda)}
V(\Lambda )\bigg]^{n}\bigg\}P(\Lambda')~.
\end{equation}
After doing the best job we can to evaluate $\cH(\Lambda')$  (it is clear one needs to truncate the infinite sum
and do something about the energy dependence of the denominators), the terms separate naturally into
a diagonal operator $H_0$ and  residual interactions $V(\Lambda')$. 
Sometimes we are lucky, and complicated terms of $V(\Lambda')$  become relatively
smaller as $\Lambda '$ is reduced. These are called {\em irrelevant interactions} which scale to zero.
We end this section by remarking that the RG transformation should preserve all 
the symmetries of the Hamitonian. If   $\cH$
has explicit symmetry-breaking terms, those terms may grow or shrink under the RG transformation
rendering the low energy correlations less or more symmetrical, as the case may be.

\section{From Hubbard to $t-J$ and Heisenberg Models}
\label{s.tJ}
As an explicit derivation of an effective Hamiltonian outlined in Sec. \ref{s.lowenproj},
We consider the  Hubbard model $\cH=\cT+\cU$ of Eq. (\ref{HM})
in the strong coupling regime ($U/t \gg 1$).
The diagonal part $H_0$ we choose as $\cU$, the onsite interactions.
This term  divides  the Fock space
 into two subspaces, the singly occupied and empty sites configurations
\begin{equation}
S = \left.  \Big\{ \left| n_{1,\uparrow},n_{1,\downarrow},
                         n_{2,\uparrow},n_{2,\downarrow},\ldots \right>
                         / \forall i, n_{i,\uparrow}+n_{i,\downarrow}
                         \leq 1 \Big\}~, \right.
\end{equation}
and configurations with one or more
doubly occupied sites
\begin{equation}
D = \left. \Big\{ \left| n_{1,\uparrow},n_{1,\downarrow},
                         n_{2,\uparrow},n_{2,\downarrow},\ldots \right>
                         / \exists i, n_{i,\uparrow}+n_{i,\downarrow}
                         = 2  \Big\}~. \right.
\end{equation}
The hopping term $\cT$  couples the $S$ and $D$ subspaces by moving an electron into, or out
of, 
a doubly occupied state. We define $P_0$ to project onto the ground state manifold of
subspace $S$, and thus 
\begin{equation}
G_{00}(E)=P_{0}G(E)P_{0}
=\left(E - {\mathcal H}_{eff}(E)\right)^{-1}~,
\end{equation}
where the effective Hamiltonian ${\mathcal H}_{eff}$, is given by
Eq.~\refq{e.effham}
\begin{equation}
{\mathcal H}_{eff}(E) = P_{0}{\mathcal T}P_{0} + 
P_{0}{\mathcal T} \Big[ (1-P_{0})(E-({\mathcal U + T}))(1-P_{0}) \Big]^{-1}
{\mathcal T}P_{0}~.
\end{equation}
In the strong coupling limit, expanding the effective Hamiltonian
to zeroth order in $E/U$ and to second order in $t/U$ one gets
\begin{eqnarray}
{\mathcal H}_{eff}(E) &\xrightarrow [t/U \ll 1]{}& {\mathcal H}^{t-J}~,\\
{\mathcal H}^{t-J} &=&  P_{0}\left( {\mathcal T} - \frac{t^{2}}{U}
\sum_{i,j,k,s,s^{\prime}}
c_{i,s}^{\dagger}c_{j,s}
n_{j,\uparrow}n_{j,\downarrow}
c_{j,s^{\prime}}^{\dagger}c_{k,s^{\prime}}
\right)P_{0}~,
\label{e.tjmodel}
\end{eqnarray}
i.e., the low energy excitations of the Hubbard model
are described by the Hamiltonian of the so  called $t-J$ model.

The fermion operators appearing in Eq.~\refq{e.tjmodel}, 
can be rearranged in the following way:
\begin{eqnarray}
{\mathcal H}^{t-J} & = & P_{0}\left({\mathcal T} + {\mathcal T}^{\prime} 
                        + {\mathcal H}^{H}\right)P_{0}~, \\
{\mathcal T}^{\prime} & = &  - \frac{t^{2}}{2U}\sum_{i,j,k}^{i\neq k}\,
\left[\sum_{s}c_{i,s}^{\dagger}c_{k,s}n_{j}
-c_{i}^{\dagger}\vec{\sigma}c_{k}\cdot c_{j}^{\dagger}\vec{\sigma}c_{j}\right]~, \\
{\mathcal H}^{H} & = & \frac{J}{2}\sum_{\ij}\left({\bf S}_{i}\cdot{\bf S}_{j}
-\frac{n_{i}n_{j}}{4}\right)~,
\end{eqnarray}
where $J=4t^2/U$ and the $S=1/2$ spin operators ${\bf S}_{i}$ are 
\begin{equation}
S_{i}^{\alpha} = \frac{1}{2}\sum_{s,s^{\prime}}\,
c_{i,s}^{\dagger}\sigma_{s,s^{\prime}}^{\alpha}c_{i,s^{\prime}}~,
\end{equation}
$\sigma^{\alpha}$ being the Pauli matrices.

At half filling, i.e., when $n_{i}=1$, $P_{0}$ annihilates
${\mathcal T}$ and ${\mathcal T}^{\prime}$ since there can be no  hopping processes within subspace $S$ when
there are no empty sites. The transport of charge is prevented by
an energy gap of order $U$. This is the  {\em Mott insulator}, which describes 
the undoped (parent compounds) of the high T$_c$ superconductors of the cuprate family.

In this limit, the t-J model simply reduces to the spin $S=1/2$ Heisenberg model
\beq
\cH^{t-J} \to  \frac{J}{2}\sum_{\ij} {\bf S}_{i}\cdot{\bf S}_{j} + \mbox{const}.
\end{equation}
As in the two site  Hubbard model of two electrons (see Sec.\ref{s.2siteHubbmod}),   
the low energy excitations are purely  magnetic. 
  
\section{The Negative-U  Hubbard Model}
\label{s.nutopu}

The negative-U Hubbard model describes local attractive
interactions between electrons which could be produced by several microscopic mechanisms
e.g.,  phonons, plasmons or spin fluctuations. We choose, for convenience to write the model as follows
\begin{multline}
{\mathcal H}^{-U} = -t\sum_{\ij,s} c_{i,s}^{\dagger}c_{j,s} 
- \frac{U}{2} \sum_{i}(n_{i}-1)^{2} + \\
+ \frac{1}{2}\sum_{\ij}V_{ij}(n_{i}-1)(n_{j}-1) - \mu\sum_{i}n_{i}~,
\end{multline}
where the negative-U term favors pairs of electrons  on the
same site in
competition with the hopping term which delocalizes the electrons;  $V_{ij}$ 
is intersite Coulomb interactions and $\mu$ the chemical potential.
The following canonical transformation
\begin{eqnarray}
c_{i,\uparrow} & \longrightarrow & \tilde{c}_{i,\uparrow}~, \\
c_{i,\downarrow} & \longrightarrow & \tilde{c}_{i,\downarrow}^{\dagger}~,
\end{eqnarray}
maps the negative-U to a positive-U Hamiltonian:
\begin{multline}
{\mathcal H}^{-U} \rightarrow {\mathcal H}^{+U} = 
-t\sum_{\ij} ( \tilde{c}_{i,\uparrow}^{\dagger}\tilde{c}_{j,\uparrow} -
\tilde{c}_{i,\downarrow}^{\dagger}\tilde{c}_{j,\downarrow} ) 
+ \frac{U}{2} \sum_{i}(\tilde{n}_{i}-1)^2 + \\
+ \frac{1}{2}\sum_{i,j}J_{ij}^a\tilde{S}_{i}^{z}\tilde{S}_{j}^{z} - h \sum_{i}\tilde{S}_{i}^{z} 
- N(\mu + \frac{U}{2})~,
\end{multline}
where
\begin{eqnarray}
J_{ij}^a &=& 4V_{ij}~, \\
h&=&2\mu~, \\
\tilde{S}_{i}^{z} &=& \frac{1}{2}(\tilde{n}_{i}-1)~,
\end{eqnarray}
and $N$ is the total number of sites.

Following the derivations of Sec. \ref{s.tJ},  and using the fact that ${\mathcal H}^{+U}$ is
at {\em half filling} (for a proof see Sec. 3.3.1 in \cite{auerbach}), this model at large $|U|/t$
can be directly mapped onto an effective pseudospin model
\bea
\cH^{+U} &\to& \cH^{-x-xz} + {\cal O} (t^2/U)\nonumber\\
\cH^{-x-xz}&=& \frac{1}{2}\sum_{ij}^{nn}\Big[J_{ij}^{z}
\tilde{S}_{i}^{z}\tilde{S}_{j}^{z}-J_{ij}^{x}(\tilde{S}_{i}^{x}\tilde{S}_{j}^{x}
+\tilde{S}_{i}^{y}\tilde{S}_{j}^{y})\Big]-\sum_{i}h_{i}\tilde{S}_{i}^{z}
\label{-x-xz}
\eea
where the {\em pseudospin} operators are
\bea
\tilde{S}_{i}^{z}&=&\frac{1}{2}(\tilde{n}_{i}-1)\nonumber\\
 \tilde{S}_{i}^{x}&=& \frac{1}{2}(\tilde{c}_{i\uparrow}^{\dagger}\tilde{c}_{i\downarrow}^{\dagger}
+\tilde{c}_{i\downarrow}\tilde{c}_{i\uparrow})\nonumber\\
\tilde{S}_{i}^{y}&=& \frac{1}{2i}(\tilde{c}_{i\uparrow}^{\dagger}\tilde{c}_{i\downarrow}^{\dagger}
-\tilde{c}_{i\downarrow}\tilde{c}_{i\uparrow})~.
\label{pseudospins}
\eea
We see that the local charge operator and the  pair operator have the
same commutation relations as angular momenta along the $z$-axis and $xy$ plane respectively.
The quantum properties of the pseudospins explains Josephson commutation 
relation between  charge and superconducting phase,
$[N,\phi]= 1$.

At weak coupling $|U|/t <1$, it can be also argued that the negative-$U$ model 
renormalizes onto a similar effective model as (\ref{-x-xz}) albeit with different
lattice constant and interaction parameters. The
Fermi sea is unstable with respect to attractive interactions as seen diagrammatically 
by the divergence of
the vertex function in the BCS or charge density wave channels.
Since the attractive interaction scales to strong coupling,
at the scale where the cut-off energy equals the BCS gap,   the effective Hamiltonian
can be  transformed to the  -x-xz model  to obtain the strong coupling fixed point Hamiltonian.
This procedure however, has not yet been carried out, to the best of our knowledge.

In Lecture \ref{lec3}, we  shall use the classical $\cH^{-x-xz}$ model to
describe superconductivity,
and charge density wave phases.

\newpage


\part{Quantum Magnetism}
\label{lec2}
This lecture is technical in nature. It contains a brief review of the spin  path integral and how
to obtain the classical and semiclassical approximations to it. A fuller background
for this subject, with compatible notations,   can be found in Ref.\cite{auerbach}. Here, a new emphasis is
placed on anisotropic models and their rotator representation.
\section{Spin Coherent States}
Path integrals provide  formal expressions which can lead 
to useful approximation schemes. A path integral representation of spin models
can be constructed using {\em spin coherent states}.
Let us consider the eigenstates $\left. \left. \right|S,m\right>$ of
${\bf S}^{2}$ and $S^{z}$ with eigenvalues $S(S+1)$ and $m$, respectively. Spin coherent states
are a family of spin states labelled by a unit vector $\hat{\Omega}=(\theta,\phi)$, where $\theta$ and $\phi$ are
the lattitude and longitude angles respectively.
The are defined by 
applying the SU(2) rotation operator to the highest weight state\footnote{The phase $e^{-iS\phi}$ 
represents a gauge choice
with one singularity on the sphere: at the south pole.} in  representation $S$:
\begin{eqnarray}
\big|\hat{\Omega}\big>_S &\equiv&  
e^{i\phi S^{z}} e^{i\theta S^{y}}e^{-i\phi S^{z}} \left|S,S\right>\nonumber\\
&=&e^{-iS\phi}  \sqrt{(2S)!}\sum_{m=-S}^{+S}\frac{u(\theta,\phi)^{S+m}v(\theta,\phi)^{S-m}}
{\sqrt{(S+m)!(S-m)!}} \left|S,m\right> ~,
\label{e.spcs}
\end{eqnarray}
with
\begin{eqnarray}
u(\theta,\phi) & = & \cos{(\theta/2)}e^{i\phi/2}~, \\
v(\theta,\phi) & = & \sin{(\theta/2)}e^{-i\phi/2}~. 
\end{eqnarray}
Using Eq.~\refq{e.spcs}, two useful identities can be readily proven:
\begin{itemize}
\item The resolution of  identity 
\begin{equation}
\frac{2S+1}{4\pi}\int_{-1}^{1}d\cos{\theta}\int_{0}^{2\pi}d\phi 
\left.\big|\hat{\Omega}\big>\big<\hat{\Omega}\big|\right. =  
\sum_{m=-S}^{+S}\left.\left|S,m\right>\left<S,m\right|\right.= I_{S}~.
\label{e.res}
\end{equation}
\item The overlap of two  states with closeby unit vectors
\begin{equation}
\big<\hat{\Omega}\big|\hat{\Omega}'\big>\simeq \left(\frac{1+\hat{\Omega}\cdot \hat{\Omega}'}{ 2}\right)^S
e^{\left(iS (1-\cos {\bar \theta}  )(\phi-\phi')\right)}
\label{overlap}
\end{equation}
where $ {\bar \theta}$ is the average lattitude of the two vectors. 
\end{itemize}

\section{Spin Path Integral}
\begin{figure}
\centerline{\psfig{bbllx=210pt,bblly=0pt,bburx=510pt,bbury=260pt,%
figure=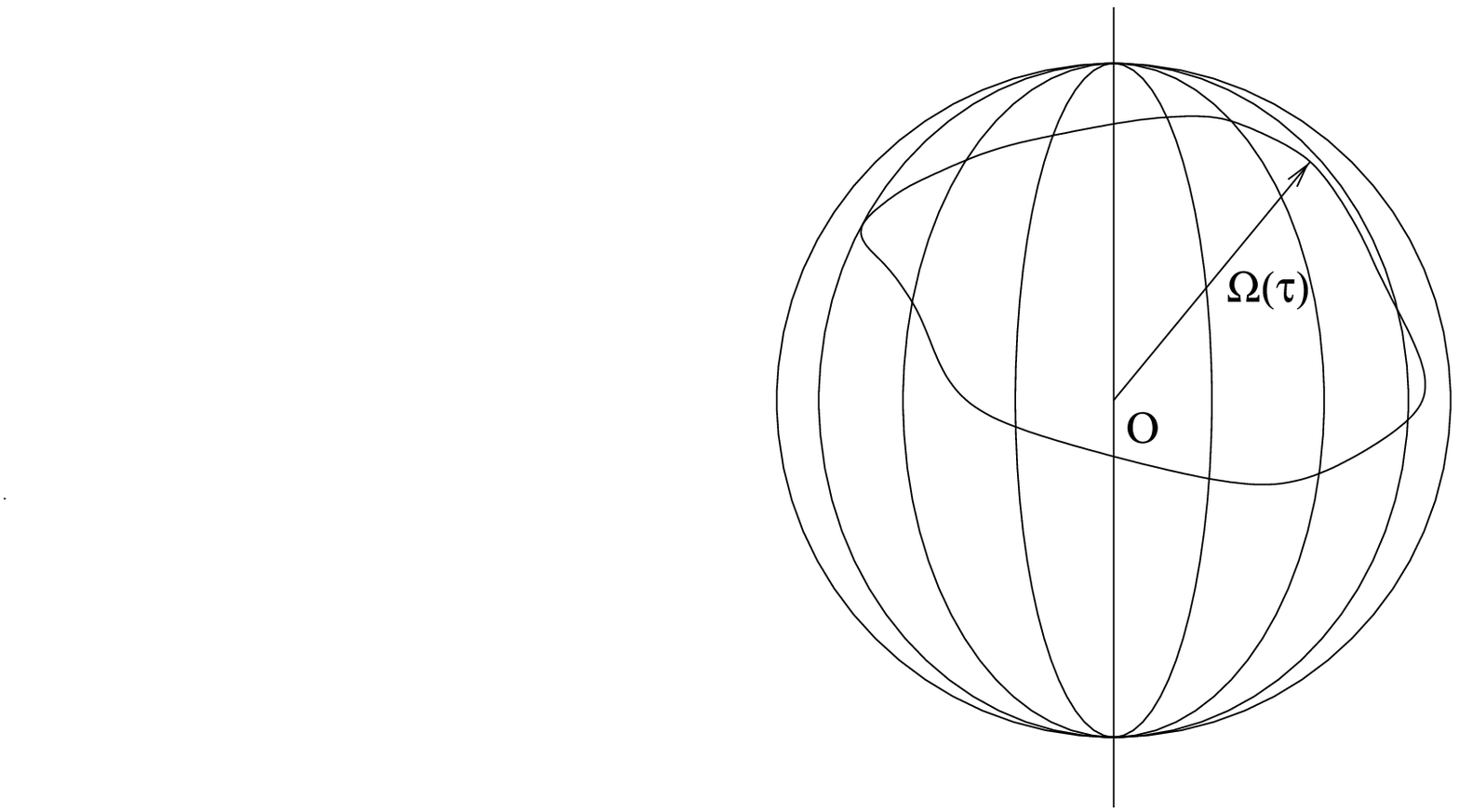,width=70mm,angle=0}}   
\caption{The Berry phase $\omega[\hat{\Omega}]$ measures the area enclosed by the 
trajectory of $\hat{\Omega}(\tau)$ on the unit sphere which does not include the south pole.}
\label{f.barry}
\end{figure}

The partition function of a single spin with Hamiltonian $H$ 
is
\begin{equation}
Z=Tr\left[e^{-\beta H}\right]=
Tr\left[\underbrace{e^{-\epsilon H}e^{-\epsilon H}\dots 
e^{-\epsilon H}}_{N_{\epsilon}=\beta/\epsilon\, {\textrm times}}\right]~,
\end{equation}
with $\beta=1/T$, $T$ being the temperature.
By inserting  $N_{\epsilon}-1$ resolutions of the identity
\refq{e.res}, and 
in the limit $\epsilon \rightarrow 0$,  
by expanding each exponential to  first order, one gets
\begin{equation}
Z\simeq\int d\hat{\Omega}_{1}\dots d\hat{\Omega}_{N_{\epsilon}}
\prod_{n=0}^{N_{\epsilon}-1}
\big<\hat{\Omega}(\tau_n)\big|1-\epsilon H\big|\hat{\Omega}(\tau_{n+1})\big>~,
\label{e.partfunc}
\end{equation}
with the boundary condition $\hat{\Omega}_0=\hat{\Omega}_{N_{\epsilon}} $.
In the limit $N_{\epsilon}\rightarrow \infty$ (and $\epsilon\rightarrow 0, 
\beta=N_{\epsilon}\epsilon=
 \textrm{const}$.) Eq.~\refq{e.partfunc} defines a path integral
\begin{equation}
Z=\oint{\mathcal D}\hat{\Omega}(\tau) \exp\left[-\int_{0}^{\beta}
d\tau\left(iS(1-\cos{\theta(\tau)})\dot{\phi}(\tau)+H\big[\hat{\Omega}(\tau)\big]\right)\right]~,
\label{e.pathintz}
\end{equation}
The time dependent term in Eq.~\refq{e.pathintz} 
\begin{equation}
iS\omega\left[\hat{\Omega}\right] = iS \int_0^\beta  d\tau (1- \cos\theta){\dot  \phi}
\label{berry}
\end{equation}
derives from the  overlap between coherent states (\ref{overlap}). It
is known as the {\em Berry phase} of the spin history and it is  
geometric, i.e.,
depends on the trajectory of $\hat{\Omega}(\tau)$ on the unit sphere. In fact it measures the area enclosed
by the path $\hat{\Omega}(\tau)$ on the unit sphere (Fig.~\ref{f.barry}).

\vspace{0.5cm}
The classical Hamiltonian is defined as
\begin{equation}
H\big[\hat{\Omega}(\tau)\big]=\big<\hat{\Omega}(\tau)\big|H\big|\hat{\Omega}(\tau)\big>~.
\end{equation}
An implicit assumption  in Eq. \refq{e.pathintz} is that 
the path integral is dominated by smooth (differentiable) paths. This turns 
out to be unjustified, since  discontinuos paths matter for the 
correct ordering of  quantum operators. For that reason, path integral 
results should be checked whenever possible against operator methods. Ordering ambiguities give
rise to erroneous quantum corrections to  energies and local spin correlations. 
They do not effect, however,
long distance and long timescale correlation functions.

The partition function for a system of $N$ spins is
\begin{equation}
Z=\oint\prod_{i=1}^{N}{\cal D}\hat{\Omega}_{i}(\tau)
\exp{\left[
iS\sum_{i=1}^{N}\omega\big[\hat{\Omega}_{i}\big]
-\int_{0}^{\beta}d\tau
H\big[\{\hat{\Omega}_{i}(\tau)\}\big]\right]}~.
\label{e.spcohzeta}
\end{equation}
For example, the nearest neighbor Heisenberg model partition function is  
\begin{equation}
Z_{H}=\oint{\cal D}\hat{\Omega}\exp{\left[iS\sum_{i=1}^{N}
\omega\big[\hat{\Omega}_{i}\big] - \frac{J}{2}\int_{0}^{\beta}
\sum_{\ij}\hat{\Omega}_{i}\cdot\hat{\Omega}_{j}\right] }~.
\label{nnHM}
\end{equation}

The spin coherent states path integral \refq{e.spcohzeta}
are
convenient starting points for deriving semiclassical, i.e. large $S$, approximations.
The integration variables are unit vectors, i.e.  classical spins.
The quantum effects enter through their time dependent fluctuations. Keeping $JS^2\to J$ fixed
and sending
$S\rightarrow\infty$, suppresses the contributions of  fluctuating paths with $\dot{\hat{\Omega}}\neq0$.
This leaves integration over frozen spin configurations precisely as in the classical partition function
\begin{equation}
Z\xrightarrow[S \rightarrow \infty]{} \int \prod_{i=1}^{N} d\hat{\Omega}_{i}
e^{-\beta H\left[\{\hat{\Omega}_{i}\}\right]}~.
\end{equation}
Now it is possible to use $S$ as the control parameter for a
systematic expansion of the partition  function. In particular,
applying the saddle point approximation (analytically continued to {\em real} time $t=i\tau$)  yields  
\begin{equation}
S\frac{\delta\omega\big[\hat{\Omega}_{i}\big]}{\delta\hat{\Omega}_{i}}
-\int_{0}^{t}dt\frac{\delta H}{\delta\hat{\Omega}_{i}}=0~,
\end{equation}
which are the {\em Euler-Lagrange} equations of motion for   classical spins
\begin{equation}
S\hat{\Omega}_{i}^{cl}(t)\times\dot{\hat{\Omega}}_{i}^{cl}(t)=
\frac{\partial
}{\partial\hat{\Omega}_{i}}H\left[\{\hat{\Omega}^{cl}_{j}\}\right]~.
\end{equation}

\section{Spin Wave Theory}
When spin symmetry is broken, either spontanously or by explicit symmetry breaking terms, 
it is quite natural to use the semiclassical expansion of the path
integral. We consider
small fluctuations around a classical spin configuration, $\hat{\Omega}_{i}^{cl}$ which minimizes $H[\Omega]$.
\begin{equation}
\hat{\Omega}_{i}(\tau)=\hat{\Omega}_{i}^{cl}+\delta\hat{\Omega}_{i}(\tau)~.
\end{equation}
To leading gaussian order, the partition function is approximated by
\begin{equation}
Z\simeq e^{-\beta H\left[\Omega^{cl}\right]}
\int{\cal D}\delta\hat{\Omega}(\tau)
\exp{\left[iS\sum_{i=1}^{N}\delta^{2}\omega\big[\hat{\Omega}_{i}\big]
-\int_{0}^{\beta}d\tau
\delta^{2}H\big[\{\hat{\Omega}_{i}(\tau)\}\big]\right]}~,
\end{equation}
where
\begin{eqnarray}
\delta^{2}\omega[\hat{\Omega}_{i}\big]&=&\frac{1}{2}\sum_{i=1}^{N}\int_{0}^{\beta}d\tau\,
\hat{\Omega}_{i}\cdot\delta\dot{\hat{\Omega}}_{i}\times\delta\hat{\Omega}_{i}~,\nonumber\\
\delta^{2}H[\hat{\Omega}_{i}\big]&=&\frac{1}{2}\sum_{\ij}
\delta\hat{\Omega}_{i}\frac{\delta^{2}H[\{\hat{\Omega}_{i}\}\big]}
{\delta\hat{\Omega}_{i}\delta\hat{\Omega}_{j}}\delta\hat{\Omega}_{j}~.
\end{eqnarray}
$\delta\hat{\Omega}_{i}$, which are perpendicular to $ \hat{\Omega}_{i}$  
can be projected onto the two tangential unit vectors which defines the harmonic oscillator degrees of freedom
\begin{eqnarray}
q_{i} &=& \delta\hat{\Omega}_{i}\cdot \hat{\phi}_{i} \\
p_{i} &=& S\delta\hat{\Omega}_{i}\cdot \hat{\theta}_{i}~,
\end{eqnarray}
These variables can be used to represent the gaussian fluctuations as
\begin{equation}
Z\simeq e^{-\beta H\left[\hat{\Omega}^{cl}\right] }
\int{\cal D}p_{i}{\cal D}q_{i}\exp{\left[\int_{0}^{\beta}d\tau\left(
i\frac{{\bf p}\cdot\dot{\bf q}-\dot{\bf p}\cdot{\bf q}}{2} 
- \frac{1}{2} \left( {\bf q},{\bf p} \right) H^{(2)}
\left( \begin{array}{c} {\bf q} \\ {\bf p} \end{array}\right)\right)\right]}~,
\label{SWT}
\end{equation}
where $H^{(2)}$ is a dynamical matrix of coupled harmonic oscillators
\begin{equation}
H^{(2)}=\left(\begin{array}{cc} 
K     &   P    \\
P^{t} & M^{-1}
\end{array}\right)~,
\end{equation}
where
\begin{eqnarray}
K_{ij} & = & \left.\frac{\partial^{2} H}{\partial q_{i} \partial  q_{j}}
\right|_{{\bf q}={\bf p}=0}~, \\
M^{-1}_{ij} & = & \left.\frac{\partial^{2} H}{\partial  p_{i} \partial
p_{j} }
\right|_{{\bf q}={\bf p}=0}~, 
\end{eqnarray}
are the force constant and reciprocal mass matrices respectively and
\begin{equation}
P_{ij}= \left.\frac{\partial^{2} H}{\partial p_{i}\partial  q_{j}}
\right|_{{\bf q}={\bf p}=0}~, 
\end{equation}
couples coordinates and momenta.
Eq. (\ref{SWT}) is  the harmonic spin wave partition function of any quantum spin Hamiltonian, 
whose classical ground state is known. By diagonalizing its action one readily obtains the spin wave excitation
energies and wavefunctions, and spin correlations can be evaluated to the subleading order in $S^{-1}$. 
The complexity of the calculation depends on the lattice symmetry of the classical ground state, i.e.
such as the size of its {\em magnetic}  unit cell.  For example, for the  N\'eel state, it is two lattice unit cells.

For completeness, we work out the spin wave dispersion of the antiferromagnetic Heisenberg model
with a N\'eel state given by
\begin{equation}
\left( \theta_{i}^{cl},\phi_{i}^{cl}\right) = \left\{
\begin{array}{cc}
\left(\frac{\pi}{2},0\right) & i\in{\textrm A} \\
\left(\frac{\pi}{2},\pi\right) & i\in{\textrm B} 
\end{array} \right.
\end{equation}
where A and B are the two sublattices in which the lattice can be divided.
The  harmonic degrees of freedom are
\begin{equation}
q_{i}(t) =\left\{
\begin{array}{cc}
\phi_{i}(t) & i\in{\textrm A} \\
\pi + \phi_{i}(t) & i\in{\textrm B} 
\end{array} \right.
\end{equation}
and
\begin{equation}
p_{i}(t)=S\cos{\theta_{i}(t)}~.
\end{equation}
The dynamical matrix of the model (\ref{nnHM}) is
\begin{multline}
H^{(2)} = \frac{J}{2}\sum_{\ij}
\left[\frac{(p_{i}-p_{j})^2}{S^{2}} - (q_{i}-q_{j})^{2}\right] = \\
=\frac{1}{2}\sum_{\bk} \left(p_{\bk},q_{\bk}\right)
\left(\begin{array}{cc} zJ(1+\gamma_{\bk}) & 0 \\
0 & zJS^{2}(1-\gamma_{\bk})
\end{array} \right)
\left( \begin{array}{c} 
p_{\bk} \\ q_{\bk} \end{array} \right)~,
\end{multline}
where
\begin{equation}
\gamma_{\bk} = \frac{1}{z} \sum_{\bd} e^{i \bk \cdot {\bd}}~,
\end{equation}
$z$ and $\bd$ being respectively the coordination number
and the vector connecting one site to its nearest-neighbours.
The dispersion relation of the small fluctuations around the 
ground state configuration, i.e., of the {\em spin waves},
can be found  solving the characteristic
equation
\begin{equation}
\det \left(\begin{array}{cc} S^{-2}zJ(1+\gamma_{\bk}) & i\omega_{\bk} \\
-i\omega_{\bk} & zJ(1-\gamma_{\bk})
\end{array} \right) = 0~,
\end{equation}
and is $\omega_{\bk}=\frac{1}{ S}zJ\sqrt{1-\gamma_{\bk}^{2}}$, for two distinct spin wave modes. 

\section{Continuum Theory for Anisotropic Models}
\label{s.CONT}

Spin wave theory is restricted to the ordered
phases of the Heisenberg model.  However, one can still use a
semiclassical approach even in the absence of spontaneously
broken symmetry. A short range classical Hamiltonian is  mostly sensitive 
to short-range correlations. Thus  the Heisenberg Hamiltonian, in the
large $S$ limit, has at least short range antiferromagnetic order. 
In the path integral approach it is possible to utilize  the  short lengthscale correlations
to define a continuum theory without assuming broken symmetry.

In this section we spend some time preparing the ground for  Lecture \ref{lec3}. 
To that end,
we  derive the continuum theory
for the  anisotropic xxz model in a magnetic field. The resulting path integral 
will be later used in the context of quantum properties of superconductors. 
The continuum theory is shown to be equivalent to 
SO(3)   quantum rotators. The rotator formulation is readily generalizable
to SO(5) symmetry, which is the topic of section  \ref{s.SO5}.
Subsequently, for the isotropic case  
we review Haldane's mapping of the quantum Heisenberg antiferromagnet  in $d$ dimensions
into the nonlinear sigma model (NLSM) in $d+1$ dimensions, and the main results which can be obtained by
that mapping.

The first step is to parametrize the spins using two continuous vector fields
$\hat{n}$ and $\vec{L}$,  
\begin{equation}
\hat{\Omega}^\alpha_{i}=\eta_{i}\hat{n}^\alpha(\vec{x}_{i})\sqrt{1-\left|\frac{\vec{L}
(\vec{x}_{i})}{S}\right|^{2}}+\frac{{L}^\alpha(\vec{x}_{i})}{S}~,
\label{fields}
\end{equation}
where $\eta_{i}=e^{i\vec{\pi}\cdot\vec{x}_{i}}$ has opposite
signs on the two sublattices.
Each pair of neighboring  spins (4 degrees of freedom) is replaced by  $\hat{n},\vec{L}$.
We can choose   $\hat{n}$ to be a unimodular ($|\hat{n} |^2=1$) N\'eel field 
(2 degrees of freedom),  and $\vec{L}$ is the  
 perpendicular canting  field, with the
 constraint $\vec{L}\cdot\hat{n} =0$ (2 degrees of freedom).
The spin measure of Eq. (\ref{e.spcohzeta}) becomes
\begin{equation}
\cD \hat{\Omega} \to \cD \hat{n} \cD \vec{L} 
\delta\left( \vec{L} \cdot\hat{n} \right)
\label{measure}
\end{equation}
where the $\delta$ functionals are local space-time constraints.

Let us consider a general anisotropic spin model in a magnetic field $\vec{h}$,
 \begin{equation}
{\cal H}= \frac{1}{ 2}\sum_{ij,\alpha} J_{ij}^\alpha \hat{\Omega}^\alpha_i  \hat{\Omega}^\alpha_j - 
\vec{h}\cdot S \sum_i \hat{\Omega}_i~.
\label{+xxz}
\end{equation}
Using (\ref{fields}) and expanding to lowest order in $L^\alpha$, $\partial_i n^\alpha$ and $\partial_i L^\alpha$
 we obtain the energy density
 \begin{eqnarray}
{\cal H}&\to& \int d^d x ~ H\left[\vec{L} ,\hat{n} \right]\nonumber\\
H &=&   ~E^{cl}[n] ~+ \frac{1}{2} \sum_\alpha\left({\chi}^{-1}_{\alpha}  \left(L^\alpha \right)^2 
+  \rho_s^\alpha (\partial_i n^\alpha)^2\right)~ -  a^{-d}\vec{h}\vec{L},
\label{xxz-h}
\end{eqnarray}
where the energy, spin stiffness, and susceptibility parameters are respectively:
 \begin{eqnarray}
E^{cl}[n] &\equiv& -\frac{1}{ 2 {\cal N} a^d} \sum_\alpha ( n^\alpha)^2 \left(\sum_{ij} J_{ij}^\alpha \eta_i\eta_j\right) ,\nonumber\\
\rho^\alpha &\equiv& \frac{1}{ 2 d {\cal N} a^d}\sum_{ij} J^\alpha_{ij} \eta_i\eta_j|\vec{x}_i-\vec{x}_j|^2,\nonumber\\
\frac{1}{ \chi_{\alpha}} &=& \frac{ 2S^{-2} }{ {\cal N}  a^d} \left(\sum_{ij} J^\alpha_{ij} \right) -2S^{-2} 
E^{cl} 
\label{par}
\end{eqnarray}
$a$ and ${\cal N}$ are the lattice constant and size respectively.
Expansion of the Berry phase term to the same order yields two terms
 \begin{equation}
iS\sum_i\omega\left[\hat{\Omega}_i\right] =-i\Upsilon + i \int_0^\beta d\tau \int d^d x ~\hat{n} \times  \dot{\hat{n}}
\cdot \vec{L},
\end{equation}
where 
\begin{equation}
\Upsilon(\hat{n})=S\int d^{d}x e^{i\vec{\pi}\cdot\vec{x}}
\omega(\hat{n}(\vec{x}))~.
\end{equation}
Collecting the terms together, we have the path integral
\begin{equation}
 Z^{xxz}=\int\cD \hat{n} \cD \vec{L}\delta\left( \vec{L}\cdot\hat{n}\right)
 e^{-i\Upsilon} \exp\left( \int d\tau d^d x ~i  \hat{n} \times \dot{\hat{n}}\cdot
\vec{L}  -H[\vec{L},\hat{n}]\right).
\label{PI}
\end{equation}
\section{Anisotropic Quantum Rotators}
\label{s.ROT}
Eq. (\ref{PI}) can be physically understood as a path integral of {\em rotators}. 
Consider the phase space path integral over an $N$ dimensional field $\vec{n}$, and canonical momenta $\vec{p}$
with a ``Mexican hat'' potential
\begin{eqnarray}
Z&=&\int \cD \vec{p}\cD \vec{n}\exp\left( \int d\tau d^d x~ i
\dot{\vec{n}}\cdot {\vec{p}} - H[p,n]\right)\nonumber\\
H^{MH}&=&H^{rot}[\vec{p},\vec{n}] + K(|\vec{n}|-1)^2~.
\end{eqnarray}
If  $K$ is taken to be very large, fluctuations of
$\delta n_\parallel =|\vec{n}|-1$, and its conjugate momentum $\vec{p}_\parallel$ become 
high frequency harmonic oscillators,
which can be integrated out in the adiabatic approximation. This leaves
us with the slow degrees of freedom $\hat{n}=\vec{n}/|\vec{n}|$, and $\vec{p}_\perp$, and a renormalized Hamiltonian
\begin{equation}
Z_{K\to \infty}\propto \int \cD  \hat{n}  \cD \vec{p}\delta(\vec{p}\cdot
 \hat{n} ) \exp 
\left( \int d \tau d^d x~ i\dot{\hat{n}}\cdot \vec{p}_\perp -\tilde{H}^{rot}[\vec{p}_\perp,\hat{n}]\right)~. 
\label{rot}\end{equation}
A Faddeev-Jackiw quantization of a particle on an $N$-sphere leads to
the same constraints, as shown in Ref. \cite{BW}.

For the N=3 model,  the angular momenta and the transverse momenta are both vectors, related by
\begin{equation}
\vec{L} =\hat{n} \times \vec{p}_\perp,~~~\vec{p}_\perp =\hat{L} \times \vec{n}.
\label{L1}\end{equation}
A change of variables $\vec{p}_\perp\to \vec{L}$ has a  unit Jacobian 
\begin{equation}
\cD \vec{p}\delta(\vec{p}\cdot\hat{n}) = \cD \vec{L}\delta(\vec{L}\cdot\hat{n})~.
\label{L3} 
\end{equation}
Substituting (\ref{L1}) and (\ref{L3}) into (\ref{rot}) yields a path integral of the form (\ref{PI}),
without the Berry phase $e^{-i\Upsilon}$.

For general $N$,
the  $N(N-1)/2$ angular momenta are defined as
\begin{equation}
L_{ab}\equiv n_a p_b - n_b p_a,~~~a,b=1,\ldots, N.
\label{L2}
\end{equation}

Note, that it is not useful for $N>3$ to write the path integral measure in terms of $L_{ab}$, 
because they are not independent
degrees of freedom, and more constraints are required.
On the other hand, the {\em transverse momenta}, which obey $\sum_a n_a (p_\perp)_a =0$, 
can be expressed as
\begin{equation}
(p_\perp)_a = \sum_b L_{ab} n_b.
\end{equation}
Therefore it  is always possible
to  express the  rotator Hamiltonian $\tilde{H}^{rot}$ of (\ref{rot}) in terms of $L_{ab}$ and $\hat{n}$. 

The opposite direction, might be even more useful.  
For example, as in Eq. (\ref{xxz-h}), the starting point could be a
Hamiltonian whose kinetic energy is expressed using symmetry generators 
 \begin{eqnarray}
H^{rot}[\vec{L},\hat{n}] &=&   ~E^{cl}[\hat{n}] ~+ \frac{1}{ 2} \sum_{a<b}  {\chi}^{-1}_{ab} L_{ab}^2
+   \frac{1}{ 2} \sum_{a} \rho_s^a (\partial_i n_a)^2 \nonumber\\
&&~~~~~~~~~~~~~~~~~~~~~~ -  a^{-d}\sum_{a<b} h_{ab}L_{ab},
\label{xxz-h-N}
\end{eqnarray}
where $h_{ab}$ and $\chi_{ab}$ are SO(N)   fields and
susceptibilities respectively. (For SO(3) their vector notation
is given by  $X^{a}\equiv \sum_{bc}\epsilon^{abc}X_{bc}$). 
Substituting (\ref{L2}) into $ H^{rot}$ yields
 \begin{eqnarray}
H^{rot}[\vec{p},\hat{n}] &=&   ~E^{cl}[n] ~+ \frac{1}{ 2} \sum_{ab}M^{-1}_{ab} p_a p_b\nonumber\\
&&~~~~~
+  \sum_a \rho_s^a (\partial_i n^a)^2~ -  a^{-d}\sum_{a<b} h_{ab} (n_a p_b-n_b p_a),\nonumber\\
M^{-1}_{ab} &\equiv& \delta_{ab}\left( \sum_{ c} \chi^{-1}_{ac} n^2_c\right) -  \chi^{-1}_{ab} n_a n_b
\label{Hrot}
\end{eqnarray}
where $M[\hat{n}]$ is an anisotropic ``mass '' matrix in the Cartesian basis.

We note that by (\ref{Hrot}), the path integral (\ref{rot}) is Gaussian in momenta $p_a$. 
One must be careful  and 
integrate only over the transverse components to $\hat{n}$. For a given direction $\hat{n}$,
we choose the transverse basis
$\hat{e}_i$, $i=1,N-1$ 
which obeys  the following conditions
\begin{eqnarray}
\hat{e}_i \cdot \hat{n}&=&0, ~~~ i=1,N-1\nonumber\\
\sum_{a,b}^N \hat{e}^a_i M^{-1}_{ab}  \hat{e}^b_j &=&\delta_{ij} {\tilde M}^{-1}_{i}[\hat{n}]~.
\label{cond}
\end{eqnarray}
This is always possible since the first condition leaves the freedom to perform an $\mbox{SO(N-1)}$ rotation
on the transverse basis. For an arbitrary  transverse basis $\{\hat{f}_i\}$,  we find the rotation  
which diagonalizes $ \hat{f}_i {M}^{-1} \hat{f}_j $, and the resulting
eigenbasis is chosen as $\{\hat{e}_i\}$.

Thus, we parametrize
\begin{equation}
\vec{p} =\sum_i p_i \hat{e}_i
\end{equation}
and integrate unrestrictedly over $\int\cD p_i$ to obtain
\begin{eqnarray}
Z&=& \int  \cD \hat{n} 
\exp\Bigg(- \int_{0}^{\beta} d\tau \int 
d^{d}x\Big(   E^{cl}[n] + \sum_{l,a}
\frac{1}{ 2}\rho^\alpha(\partial_{l}n^a)^{2}\nonumber\\
&&~~~~~~~~~~~~~~~~~~
+\frac{1}{ 2}\sum_{j} M_j
\left(\dot{\hat{n}}\cdot \hat{e}_j ~- i h_j) \right)^2
\Big)\Bigg) \nonumber\\
h_j &\equiv& a^{-d} \sum_{a<b}h_{ab}(n_b \hat{e}^a_j-n_a \hat{e}^b_j)\label{anis}~.
\end{eqnarray}
This expression is ready for the evaluation
of the classical (time independent)  ground state $\hat{n}^{cl}$ as a function of applied field $h_{ab}$,
and a spinwave expansion about it.
In the next section  however, we shall  only deal with the isotropic case.

\section{Haldane's Mapping}
We now return to (\ref{anis}) but specialize to the {\em isotropic} Heisenberg model without
a field,
where life simplifies considerably. 

The inverse mass matrix is simply
\begin{equation}
M^{-1} = \chi^{-1} \delta_{ab}- n_a n_b~.
\end{equation}
Any choice of transverse basis $\hat{e}_i$ yields a diagonal $M_i = \chi$ in Eq. (\ref{cond}).
Omitting the constant $E^{cl}$, and inserting the Berry phase $e^{-i\Upsilon[\hat{n}]}$
the partition function (\ref{anis}) reduces to  Haldane's result 
\begin{eqnarray}
Z&=&\int  \cD \hat{n} e^{-i\Upsilon(\hat{n})}\exp{\left(-\frac{1}{2}\int_{0}^{\beta} d\tau \int 
d^{d}x(\chi |\dot{\hat{n}}|^{2}+\rho_{S}\sum_{l=1}^{d}
|\partial_{l}\hat{n}|^{2})\right)}=\nonumber\\
&=&\int \cD  \hat{n} e^{-i\Upsilon(\hat{n})}\exp{\left(-\frac{a^{1-d}}{2f}\int dx_{0} 
d^{d}x~(\partial_{\mu}\hat{n})^{2}\right)}~,
\label{pha}
\end{eqnarray}
where $x_{0}=c\tau$, and $c=\sqrt{\rho_{S}/\chi}$ is the {\em spin wave
velocity}. For the nearest neighbor model with interaction $J$, one obtains from (\ref{par}):
\begin{eqnarray}
\chi&=&S^2/(4dJa^{d})~,\\
\rho_{S}&=&J a^{2-d}~,\\
f&=&\frac{ca^{1-d}}{\rho_{S}}~=2\sqrt{d}S^{-1}~.
\end{eqnarray}
Eq.~(\ref{pha}) is the partition function for a NLSM with an additional
Berry phase term.

In $1+1$ dimensions the NLSM is disordered for all $f$, as required by the classical Mermin Wagner theorem.
Its correlations are known to fall off exponentially at large distances with a correlation length which
goes as $\xi \propto e^{2\pi/f}$. By the (Lorentz) symmetry of the action  between  spatial and temporal 
dimensions, this implies a gap ({\em Haldane's gap}) for all excitations above the ground state. However
 one should also consider
the effects of the phases brought about by the
term $\Upsilon(\hat{n})$.

For $d=1$,  $\Upsilon(\hat{n})$ is a topological winding number of the two dimensional
NLSM. For all continuous fields, it yields
$e^{-i\Upsilon(\hat{n})}=e^{-i2\pi S k}$ with $k$ an integer number. Thus,
the Berry phase factor is $1$ for all integer S, while
it can be $\pm 1$ for half integer spins. As a result, it
 produces interference effects for half odd integer spins, and drastically changes the ground state 
properties and
the elementary excitations spectrum of the Heisenberg chain.

In $d = 2$
the topological phase is zero for all continuous fields.
For the nearest neighbours Heisenberg model, Neves and Perez \cite{nevper} proved that
the 
ground state is ordered for all $S \ge 1$. Also, series expansions and numerical
simulations provide evidence of the presence of an ordered ground state even for
$S=1/2$.

\section{Spin Liquid States}

For most antiferromagnetic
Heisenberg models the ground state is not explicitly known. While for finite bipartite lattices, ground state
theorems require the ground state to be a total singlet, and have positivity conditions (Marshall's signs).
The discussion of the previous section  expects them to exhibit long range order in two dimensions.
Considering these conditions, particularly useful variational states for the $S=\frac{1}{2}$ antiferromagnetic
Heisenberg model are Fazekas and Anderson's  {\em Resonating Valence Bond
} (RVB) states \cite{anderson}
\begin{equation}
\left|\{d_{\alpha}\}\right>=\sum_{\alpha}d_{\alpha}\left|\alpha\right>~,
\label{RVB}
\end{equation}
where 
\begin{equation}
\left|\alpha\right> =\prod_{(i,j)\epsilon
\Lambda_{\alpha}}^{i\epsilon A,j\epsilon B}
\frac{1}{\sqrt{2}}\Big(\left|\uparrow _{i}\right>\left|\downarrow _{j}\right>
-\left|\downarrow_{i}\right>\left|\uparrow_{j}\right>\Big)~,
\end{equation}
and
$d_{\alpha}$   can be chosen to
have the form
\begin{equation}
d_{\alpha} = \left(\prod_{(i,j)\epsilon
\Lambda_{\alpha}}u_{ij}\right)
~.
\end{equation}

Since the states $\left|\alpha\right>$ are not orthogonal to each other,
it is not possible to evaluate correlations of RVB states analytically.
Monte Carlo
simulations of Liang, Doucot and Anderson\cite{LDA}, and Havilio\cite{Moshe},  have
found that the RVB states have 
long range N\'eel order  for $u_{ij}$ that decay slower than
\begin{equation}
u_{ij}\simeq|x_{i}-x_{j}|^{-p}\,~,\,  p\leq 3~.
\end{equation}
The RVB states Eq.(\ref{RVB}) can thus be used as variational ground states for
both ordered and disordered phases. This makes them appealing
candidates for studying the transitions from the N\'eel phase
to possible quantum disordered phases, particularly in the presence of hole doping\cite{Moshe}.

\newpage
\part{Pseudospins and Superconductivity}
\label{lec3}
The -x-xz model, encountered in  Lecture \ref{lec1}, will be considered 
as an effective 
Hamiltonian for the low temperature, long wavelength properties
of $s$-wave superconductors and charge density waves. Such transitions are observed in e.g. doped
bismuthates Ba$_{1-x}$K$_x$BiBO$_3$. Recall Eq. (\ref{-x-xz}):
\begin{eqnarray}
\cH^{-x-xz}&=& \frac{1}{2}\sum_{ij}^{nn}\Big[J_{ij}^{z}
\tilde{S}_{i}^{z}\tilde{S}_{j}^{z}-J_{ij}^{x}(\tilde{S}_{i}^{x}\tilde{S}_{j}^{x}
+\tilde{S}_{i}^{y}\tilde{S}_{j}^{y})\Big]-\sum_{i}h_{i}\tilde{S}_{i}^{z}.
\label{-x-xz1}
\end{eqnarray}
$\tilde{\bf S}$ are spin $1/2$ operators, and $h=2\mu$, where $\mu$ is the
electron
chemical potential. As shown later, it can also be used to describe superfluids.
In this lecture we use the continuum rotator model, and its  classical limit
to obtain  the phase diagram.
The generalization to Zhang's SO(5) rotators, which describe the collective modes of
high T$_c$ cuprates, and the
transition from antiferromagnetism to $d$-wave superconductivity,
is straightforward. The last section uses the -x-xz model  to describe dynamics of vortices in 
superfluids.

\section{Charge Density Wave to Superconductivity}
\label{s.SO3}
In {\em bipartite lattice} with sublattices $A$ and $B$ (e.g. square, simple cubic, etc.),
the negative $J^x$ terms of the -x-xz model can be
rotated to positive terms by a global $e^{iS^z \pi}$  rotation on sublattice $B$. 
Obviously, for frustrated (non bipartite)  lattices the xxz  and the -x-xz models are {\em not} 
equivalent. For example, the triangular, Kagom\'e, and face centered cubic lattices\cite{MAA},
the -x-xz model prefers to order  in the $xy$ plane even when $J^z > |J^x|$ at zero 
doping\footnote{This helps to explain  superconductivity in 
K$_3$C$_{60}$ which
is an FCC compound with a half filled conduction band.}.  

Here we specialize to the classical model on a bipartite (square or cubic)  lattice
\begin{eqnarray}
H&=&\frac{1}{z}\sum_{\ij}^{nn}\left[ J^z \hat{\Omega}_{i}^{z}
\hat{\Omega}_{j}^{z}-J^x( \hat{\Omega}_{i}^{x}  \hat{\Omega}_{j}^{x}+
 \hat{\Omega}_{i}^{y} \hat{\Omega}_{j}^{y})\right]- \nonumber \\
&&~~~~~~~~~~-h S\sum_{i} \hat{\Omega}_{i}^{z}~+H^{nnn}\nonumber \\
H_{nnn}&=&\frac{1}{z'}\sum_{i,j}^{nnn}  K \hat{\Omega}_{i}^{z}
\hat{\Omega}_{j}^{z}\nonumber\\
\label{cl-xxz}
\end{eqnarray}
where $z$ and $z'$ are 
nearest neighbor $(nn)$ and next nearest neighbor $(nnn)$ coordination numbers respectively.

By (\ref{pseudospins}) we note that the electron charge  expectation value and the superconducting
order parameters are 
\begin{eqnarray}
\langle n_i\rangle-1 &=& \hat{\Omega}^z_i \nonumber\\
\langle  {c}_{i\uparrow}^{\dagger} {c}_{i\downarrow}^{\dagger}\rangle &=& \hat{\Omega}_i^x + i\hat{\Omega}_i^y \nonumber\\
x &=&  -{\cal N}^{-1} \sum_i  \hat{\Omega}^z_i 
\end{eqnarray}
$x$ is the hole {\em doping concentration} away from half filling.

\begin{figure}
\centerline{\psfig{bbllx=310pt,bblly=0pt,bburx=0pt,bbury=410pt,%
figure=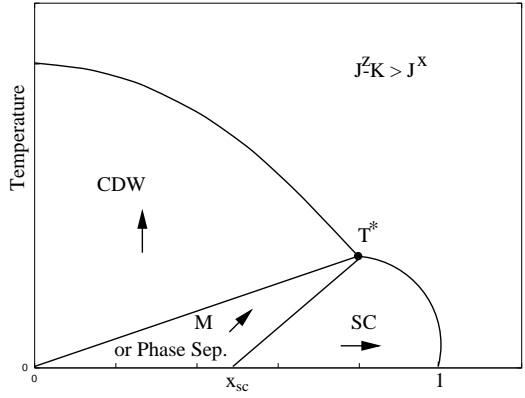,width=70mm,angle=-90}}   
\caption{Typical mean field phase diagram (doping concentration versus temperature) of the   -x-xz model.
The arrows denote directions of rotator field $\hat{n}$ in the xz plane.}
\label{fig:Tc}
\end{figure}

At finite temperatures, molecular mean field theory
for $S=1/2$ \cite{aharony} gives
the critical temperatures as a function of doping $x$ to be
\begin{eqnarray}
T_c^{CDW}(x)&=&\frac{1}{ 4} (J^z-K) (1-x^2)\nonumber\\
T_c^{SC}(x)&=& \frac{J^x x }{ 4~ \mbox{arctanh}(x) }
\label{Tc}
\end{eqnarray}
The  two curves meet at $T^*$ (See Fig. \ref{fig:Tc}). This point is tetracritical  or bicritical,
 depending on whether the transition (as a function of field $h$) is
second or
first order respectivley. In Fig. \ref{fig:Tc}  this is reflected by the nature of the
intermediate region which would be
either a mixed (``supersolid'') phase, or phase separation between pure CDW and SC domains
respectively.
The next section is devoted to determining the criteria for the order of the transition.

Near $T^*$, the lowest order expansion of the Ginzburg Landau free energy functional has SO(3) symmetry.
Thus even though the model might have high anisotropy $J^z>>J^x$, near the bicritical point
the Heisenberg symmetry is approximately restored. This  ``symmetry restoring'' 
is also argued to happen near
the multicritical point of the anisotropic SO(5) model\cite{Zhang}.

\section{Order of Transition from Rotator Theory}
\label{order-rot}
Finding out the  order of the CDW-SC transition at zero $T$, requires the energy of the  putative mixed state (M),  
which interpolates between the pure CDW and 
pure SC. 
The rotator partition 
function (\ref{anis}) comes in handy for that purpose.  
We write (\ref{cl-xxz}) as (\ref{+xxz}) by letting $-J^x\to +J^x$, and take the continuum  limit
with unit lattice constant $a=1$ following the perscriptions of section \ref{s.CONT}. 
The M ground state is parametrized by one N\'eel angle $\hat{n}=(\sin\theta,\cos\theta)$, where $0<\theta<\pi/2$.
$\theta=0$, $\theta=\pi/2$ are the pure CDW, SC  respectively.

First we evaluate the  inverse mass matrix of (\ref{Hrot})
\begin{equation}
M^{-1}_{ab}=\left(\begin{array}{ccc} 
\chi_{y}^{-1} \cos^2\theta& 0& -\chi_{y}^{-1} \cos\theta\sin\theta\\
0&\chi_{y}^{-1}\cos^2\theta +\chi_{z}^{-1} \sin^2\theta   & 0  \\
-\chi_{y}^{-1} \cos\theta\sin\theta & 0 & \chi_{y}^{-1} \sin^2\theta
\end{array}\right)~.
\end{equation}
The transverse vectors which diagonalize the projected inverse mass matrix are 
$\hat{e}_1 = \hat{y}\times\hat{n}$ and $\hat{e}_2 = \hat{y}$. The mass eigenvalues are
\begin{equation}
\tilde{M} = \left( \begin{array}{cc} 
\chi_{y} & 0\\
0 &\left( \chi_{y}^{-1} \cos^2\theta+ \chi_{z}^{-1} \sin^2\theta\right)^{-1}
\end{array} \right).
\end{equation}
Set the lattice constant to unity $a=1$, using (\ref{par}) and   (\ref{cl-xxz}) we obtain
\begin{eqnarray}
h_j&=& - h \sin\theta~\delta_{j,2}\nonumber\\
  \chi_z^{-1}(\theta) &=&\frac{2}{ S^{2}}\left((J^z+K)+(J^z-K)\cos^2\theta+|J^x|\sin^2\theta\right)\nonumber\\
 \chi_y^{-1}(\theta) &=& \frac{2}{ S^{2}}\left(|J^x|+ (J^z-K)\cos^2\theta+ |J^x|\sin^2\theta\right)~.
\end{eqnarray}
Here we have set the lattice constant to unity $a=1$.

Thus,  the classical ground state $\theta^{cl}$  minimizes the energy
\begin{eqnarray}
E^{rot}[\theta,h]&=& -(J^z-K)\cos^2\theta - J^x \sin^2\theta - \frac{1}{ 2} \sum_{j=1,2} M_j h^2_j\nonumber\\
&=& -(J^z-K)+(J^z-K-J^x)\sin^2\theta - \frac{\frac{1}{4} S^2 h^2 \sin^2\theta}{J^z-K+J^x+2K\sin^2\theta }~.\nonumber\\
\label{en-mf} 
\end{eqnarray}
A simple analysis of (\ref{en-mf}) reveals that the order of
the transition depends on the sign of $K$:

For $K<0$,  
\begin{equation}
\frac{\partial^2 E }{ \partial(\sin^2\theta)^2}\Big|_{K<0} <0,~~~~~ 0<\theta<\pi/2 
\end{equation}
i.e., there is a {\em first} order transition between $\theta^{cl}=0$ to $\theta^{cl}=\pi/2$ called a 
{\em ``spin flop''}.  This happens at a magnetic field $h_{sc}$ given by
\begin{eqnarray}
E^{rot}[0,h_{sc}]&=& E^{rot}[\pi/2,h_{sc}]\nonumber\\
\rightarrow~~~ h_{sc}   &=& 2S^{-1} \sqrt{ (J^z - K -J^x) (J^z + K +J^x)}~.
\label{hcr}
\end{eqnarray}
For any $h$, the doping concentration is given by
\begin{equation}
x(h)= 2S^{-1}   ~\frac{\partial  \mbox{min}_\theta E }{ \partial h}
\end{equation}
In the case of a first order transition, there is phase separation between the doping concentrations
of pure CDW and pure SC at  densities
\begin{equation}
x_{cdw}=0,~~~~~~~~x_{sc}= \sqrt{\frac{ J^z - K -J^x}{ J^z + K +J^x }}~.
\end{equation}

On the other hand, for $K > 0$, $E(\theta)$ is minimized at
 \begin{equation}
\sin^2\theta^{cl} =\left( \frac{1}{ 2}  h S \sqrt{ \frac{J^z-K+J^x}{ J^z-K-J^x }} -(J^z-K+J^x)\right) /2K 
\end{equation}
which indicates the existence of a  mixed phase for fields in the range
$h_{cdw} < h  <  h_{sc}$, where
 \begin{eqnarray}
h_{cdw} &=&\frac{2}{ S}\sqrt{(J^z-K+J^x)(J^z-K-J^x)}, ~~~x_{cdw}=0.
 \nonumber\\
h_{sc} &=&\frac{2}{ S}\sqrt{\frac{J^z-K-J^x}{ J^z-K+J^x}}  \left(J^z+K+J^x\right),
~~~x_{sc}= \sqrt{\frac{J^z-K-J^x }{ J^z-K+J^x}}. \nonumber\\
\end{eqnarray}

For the pure  nearest neighbor model ($K=0$), it is easy to see by (\ref{en-mf}), that
at $h_{sc}=h_{cdw}$ the mixed  state is degenerate with
a Maxwell construction of phase separation into SC and CDW. 
This degeneracy is  lifted in the quantum version of the same model, which favors  phase separation\cite{ehud}.

\section{SO(5) Rotator Theory and High-T$_{c}$ Superconductors}
\label{s.SO5}

Recently an  SO(5) theory of high T$_c$ cuprate superconductors has been proposed by
Shou-Cheng Zhang \cite{Zhang}. The order parameters of antiferromagnetism (AFM) and $d$-wave 
superconductivity (dSC) are written as
the cartesian components
of a 5 dimensional vector
\begin{eqnarray}
\hat{n}&=&(n_{1},n_{2},n_{3},n_{4},n_{5}), ~~~~~|\hat{n}|=1\nonumber\\
(n_{2},n_{3},n_{4}) &=& \langle \sum_{i} e^{i\vec{\pi}\vec{x}_i}\vec{S}_{i}\rangle \nonumber\\
n_1+in_5&=& \langle \sum_{\bp} g(\bp) c^\dagger_{\bp\uparrow}  c^\dagger_{\bp\downarrow}  \rangle
\end{eqnarray}
$g(\bp)$  is the Fourier transform of the pair wavefunction
which for $d$-wave pairing transforms as  $ \cos(p_{x})-\cos(p_{y})$ under
lattice rotations. Zhang has found  explicit, second quantized constructions 
for the 10 SO(5) generators $\{L_{a,b}\}_{a b}$, $a,b=1,\ldots,5 $ 
which rotate $\hat{n}$ in 5 dimensions. Particularly interesting are the $\Pi$ operators, i.e.
generators $L_{1,a},  a=2,3,4$, which rotate between the AFM and dSC hyperplanes. These are expected to
create new low lying pseudo-Goldstone modes near the transition.

Zhang has proposed that the long wave fluctuations
are described by an effective  rotator Hamiltonian of the form (\ref{Hrot}), 
 \begin{eqnarray}
H [L,n] &=&   \frac{1}{ 2 }   \sum_{a<b}  {\chi}_{ab}^{-1}   L_{ab}^2
+   \frac{1}{ 2}\rho \sum_{a}  |\nabla n_a|^2 \nonumber\\
&&~~~~~~~~~+ g (n_1^2 + n_5^2) - 2\mu L_{15}
\label{h-SO5}
\end{eqnarray}
where $L_{ab}$ are the generators of SO(5) algebra, and $L_{15}$ is the 
charge operator whose expectation value yields half the doping 
concentration $\langle L_{15}\rangle =x/2$.
Using the substitution (\ref{L2}), the momenta can be integrated out
of the partition function, leaving us with
\begin{eqnarray}
Z&=& \int  \cD \hat{n} 
\exp\Bigg(- \int_{0}^{\beta} d\tau \int 
d^{d}x\Big( \frac{1}{ 2}\sum_{j=1}^4 M_j
\left(\dot{\hat{n}}\cdot \hat{e}_j ~- i h_j) \right)^2\nonumber\\
&&~~~~~~~~~~~~+ \frac{1}{ 2}\rho \sum_{a=1}^5  |\nabla n_a|^2 + g (n_1^2 + n_5^2) 
\Big)\Bigg) \nonumber\\
h_j &\equiv& 2 \mu (n_5 \hat{e}^1_j-n_1 \hat{e}^5_j)~.
\label{Z-SO5}
\end{eqnarray}
We set $\dot{\hat{n}} =0$, and search for the classical ground state.

Let us first consider the SO(5) symmetric model, where all  $\chi_{ab}=\chi$, and $g=0$. For any
finite $\mu\ne 0$, the ground state flops into $(n_1,n_5)$ plane, i.e. is superconducting.

Experimentally, a transition from AFM to dSC is observed at low hole concentrations $x$ in many
of the high T$_c$ systems. Appealing to the analogy with the  -x-xz model,  
this suggests that the symmetry breaking terms of $SO(5)  \to SO(3) \times SO(2)$
should not be forbiddingly large.
Symmetry breaking terms can be included by  $g>0$,  and letting the charge susceptibility $\chi_c=\chi_{1,5}$
be different than all other susceptibilities $\chi_{ab}=\chi, a,b\ne1,5$.  
Without loss of generality, we can choose
the order parameter to tilt between the AFM and dSC hyperplanes
$\hat{n}=(\sin\theta,\cos\theta,0,0,0)$.
Following the same derivation as for xxz rotators(\ref{en-mf}),  the classical energy is  
\begin{equation}
E[\theta,\mu]=g \sin\theta^2 -  \frac{2\sin^2\theta  \mu^2 }{\chi^{-1}\cos^2 \theta+ 
\chi_c^{-1}\sin^2 \theta}~.
\end{equation}
It is now straightforward to verify that the ground state is in the $(n_2,n_3,n_4)$ sphere
at $\mu=0$, and will ``flop'' into the SC state $\theta=\pi/2$ at large enough $\mu$.

It also follows, using the same path as in Section \ref{order-rot},
that the order of the transition depends 
on the relative magnitudes of  susceptibilities\footnote{These results
differ somewhat from the phase boundaries in Ref.\cite{Zhang}}:
\begin{eqnarray}
 \chi_c > \chi   ~&\Rightarrow& ~~\mbox{First order transition at}~~ \mu_{dsc}=\sqrt{ g/(2 \chi_c)}\nonumber\\
&& ~~~\mbox{phase separation at} ~~0 < x < 4 \sqrt{g \chi_c/2}\nonumber\\
\chi_c < \chi   ~&\Rightarrow& ~~\mbox{Mixed phase for }~~ \sqrt{ g/(2 \chi)} < \mu <\sqrt{ g\chi /2}\left(\chi_c^{-1}
+ \chi^{-1} \right)\nonumber\\
&& ~~ ~\mbox{at densities} ~~0 < x < 4 \sqrt{g \chi/2}\nonumber\\
\end{eqnarray}
In the mixed phase, the relation between the SC order parameter and the doping concentration is linear
\begin{equation}
\langle n_1^2 + n_5 ^2 \rangle =\sin^2\theta^{cl}= \frac{x }{ 4 \sqrt{g \chi/2} }
\end{equation}
Before attempting to compare these results to experiments, we must
remember that this is  merely the classical approximation.

\section{Vortex Dynamics in Superfluids}

We heneceforth restrict ourselves to
two dimensions.
The quantum -x-xz Hamiltonian (\ref{-x-xz}) for spin $1/2$ 
can be written in terms of Holstein-Primakoff bosons defined as
\begin{eqnarray}
S^{z}=a^{\dagger}a-\frac{1}{2}~,\\
S^{-}=\sqrt{1-a^{\dagger}a}\,a~,\\
S^{+}=a^{\dagger}\,\sqrt{1-a^{\dagger}a}~,
\label{HP}
\end{eqnarray}
yielding
\begin{multline}
{\mathcal H}^{-x-xz}=\frac{1}{2}\sum_{ij}J_{ij}^{z}(\frac{1}{2}-a_{i}^{\dagger}a_{i})
(\frac{1}{2}-a_{j}^{\dagger}a_{j}) - \\ 
-J_{ij}^{x}
\big[a_{i}^{\dagger}a_{j}\sqrt{(1-a_{i}^{\dagger}
a_{i})(1-a_{j}^{\dagger}a_{j})}+h.c. \big]-h
\sum_{i}(\frac{1}{2}-a_{i}^{\dagger}a_{i})~.
\label{latt-mod}
\end{multline}
The partition function can be written as a Bose coherent state path integral
\begin{equation}
Z=\int \cD^{2}z \exp{\left[-i \int d \tau \left(\sum_{i}z_{i}^{*}\partial_{\tau}z_{i}
-H^{-x-xz}[z_{i}^{\star}(\tau),z_{i}(\tau)]\right)\right]}~,
\end{equation}
where $\cD z_i$ is an integration over the complex plane.
Taking the continuum limit, $z_{i}\longrightarrow \phi (x_{i})$,
we can write
\begin{equation}
Z=\int \cD ^{2}\phi e^{-S[\phi]}
\end{equation}
where
\begin{equation}
S=\int d\tau d^{d}x \left[\phi^{\star}i\partial_{\tau}\phi+\frac{1}{2}a|\nabla
\phi|^{2}+V(|\phi |)\right]~,
\end{equation}
is the time dependent Ginzburg-Landau action with
\begin{equation}
V(\phi,\phi^*)=zJ^{z}a^{2}\left(\frac{1}{2}-|\phi|^{2}\right)^{2}-zJ^{x}a^{2}|\phi|^{2}
\left(1-|\phi|^{2}\right)+\mu |\phi|^{2}~.
\end{equation}
The classical equation of motion is given by analytically continuing $\tau\to it$ and finding the saddle point
\begin{equation}
\frac{\delta S}{\delta \phi^{\star}}(\phi)=0 \rightarrow
i\partial_{t}\phi=a\nabla^{2}\phi +\frac{\delta V}{\delta \phi^{\star}}
(\phi ,\phi^{\star})~,
\label{edm}
\end{equation}
which is known as the Gross-Pitaevski or Non Linear Schrodinger equation,
whose solutions $\phi(x,t)$ describe  collective modes (phase fluctuations),
and  dynamics of vortex configurations.

A different approach to dynamics of superfluids, is  to  use the  quantum spin model (\ref{latt-mod})
on a lattice with a lattice spacing $a$ which is smaller than the interparticle spacing.
as represented by the spin
coherent states path integral (\ref{e.spcohzeta}).  
\begin{equation}
Z=\int \cD  \cos \theta_{i} \cD \phi_{i} \exp{\left[\int_{0}^{\beta} d\tau
\left(i\sum_{i}(1-\cos{\theta_{i}(\tau)})\dot{\phi}_{i}(\tau)-H[\hat{\Omega}_{i}]\right)\right]}~.
\end{equation}

\begin{figure}
\centerline{\psfig{bbllx=0pt,bblly=0pt,bburx=350pt,bbury=410pt,%
figure=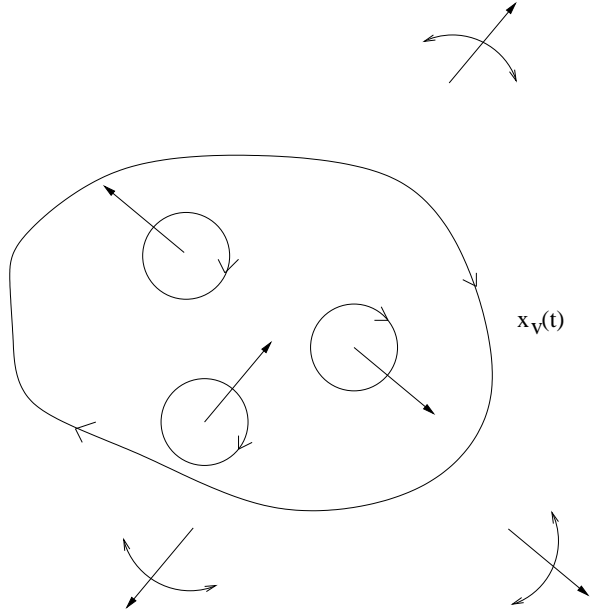,width=70mm,angle=0}}   
\caption{Berry phases due to a moving vortex.}
\label{magnus}
\end{figure} 

A superfluid state is characterized by an ordered state in the xy plane,
with a constant two dimensional boson number density $a^2 \rho_s=|\phi|^2 $. 
Using (\ref{HP}),  the average spin direction  is related to  $\rho_s$ by
\begin{equation}
\frac{1}{ 2}(1-\langle \cos{\theta}\rangle )= \rho_s a^2~. 
\end{equation}
A vortex configuration can be parametrized by  the azimuthal angles at lattice points $i$ by
\begin{equation}
\phi_{i}(t)=\arg{(x_{i}-x_{V}(t))}
\end{equation}
where $x_{V}(t)$ is the vortex core trajectory. As one can see in 
Fig.\ref{magnus},
the Berry phase of a vortex path $x_{V}(t)$ can be written
\begin{equation}
\omega=\rho_{s}a^{2} \sum_{i\epsilon S_{c}}
\oint d\phi_i=2\pi \rho_{s}a^{2}N_{c}
\label{bph}
\end{equation}
where the sum is extended to the $N_{c}$ lattice sites
included by the vortex path since the contribution of the others is zero.
This Berry phase  generates a Magnus 
force on the moving vortex. This is evident when we write it as an integral over a gauge potential
\begin{equation}
\frac{\omega}{2\pi}=\int_{c}d x \vec{A}\cdot \dot{\vec{x}}=\int_{c}
\vec{A}\cdot d\vec{l}=B  S_{c}
\label{bpha}
\end{equation}
where $S_{c}$ is the area included by the path $c$, and $B$ is an effective magnetic field 
(in dimensions of  unit flux quantum) .
Thus, comparing Eq.\refq{bph} and Eq.\refq{bpha}, one can define
$B$ to be simply
\begin{equation}
B \equiv  \rho_{s} 
\end{equation}
and the Magnus force acting on the vortex can be written
\begin{equation}
\vec{F}_{Magnus}=B\hat{z} \times \dot{\vec{x}}_{V}=
 2\pi \rho_{s}  \hat{z}\times \dot{\vec{x}}_{V}~.
\end{equation}
In the absence of any other time derivative terms, the vortex moves like a massless 
particle in a strong magnetic field
restricted to the lowest Landau level.
The semiclassical momentum $\vec{p}$ of a vortex configuration
can be evaluated by computing the expectation value of the 
translation operator $T_{\vec{a}}$
\begin{eqnarray}
e^{i\vec{a}\cdot \vec{P}} &\equiv& \big<\hat{\Omega}\big|T_{\vec{a}}\big|\hat{\Omega}\big> 
=\big<\hat{\Omega}(\vec{x})\big|
\hat{\Omega}(\vec{x}+\vec{a})\big>\nonumber\\
&\simeq &
\exp{\left[iS\sum_{i}(1-\cos{\theta_{i}})\left(\phi(\vec{x}_{i})-\phi(\vec{x}_{i}
+\vec{a})\right)\right]}\nonumber\\
&\simeq & \exp{\left(i\int d^{2}x \rho_{s}  \vec{\nabla}\phi \cdot \vec{a}\right)}=
e^{i 2\pi \rho_{s} \vec{a}\cdot \hat{z}\times  \vec{x}_{V}} ~.
\end{eqnarray}
Similarly, the momentum of a vortex-antivortex pair at positions $ \vec{x}_{V},\vec{x}_{\overline{V}} $
can be computed
\begin{equation}
\vec{P}= 2\pi\rho_{s}\hat{z}\times (\vec{x}_{V}-\vec{x}_{\overline{V}})~.
\label{mvav}
\end{equation}
Since on a lattice the only distinguishable momenta are within the first Brilluoin zone, 
Eq.~\refq{mvav} implies that 
vortex-antivortex pair configurations can tunnel between different separations which belong to the
discrete family
\begin{equation}
\vec{x}_{V}'-\vec{x}_{\overline{V}}'
=  -(2\pi \rho_{s})^{-1}  \left( \vec{P}+\vec{G}\right)\times \hat{z},
\end{equation}
where $\vec{G}$ is any reciprocal lattice vector. This is precisely an
Umklapp scattering of the superfluid current by the lattice. This amounts to
quantum dissipation of the supercurrent due to continuos translation symmetry breaking of 
a lattice potential\cite{ehud}.
  
\subsection*{Acknowledgements}
A.A. thanks the organizers, in particular Prof.
Alberto Devoto, for the excellent atmosphere at Chia Laguna Summer School. 
Discussions with S.C. Zhang, E. Demler, and S. Rabello on SO(5) rotators are greatfully acknowledged.
F.B. and L.C. thank
the organizers of the {\em Workshop with Learning} for the opportunity
of active participation and for the stimulating enviroment
they were able to create. This work was partially supported by the Israel Science Foundation, 
and the Fund  for Promotion of Research at Technion. 


\end{document}